%% file: conference_101719.tex
\documentclass[conference]{IEEEtran}
\IEEEoverridecommandlockouts
\usepackage{cite}
\usepackage{stfloats}
\usepackage{amsmath,amssymb,amsfonts}
\usepackage{graphicx}
\usepackage{textcomp}
\usepackage{xcolor}
\usepackage{hyperref}
\usepackage{xcolor}
\usepackage{placeins}
\usepackage{tcolorbox}
\usepackage{listings}
\usepackage{enumitem}
\usepackage{algorithm}

\usepackage{algpseudocode}
\usepackage{amsmath}
\usepackage{amsfonts}

\def\BibTeX{{\rm B\kern-.05em{\sc i\kern-.025em b}\kern-.08em
    T\kern-.1667em\lower.7ex\hbox{E}\kern-.125emX}}
\begin{document}

\title{BRAVE: Brain-Controlled Prosthetic Arm with Voice Integration and Embodied Learning for Enhanced Mobility}

\author{
\IEEEauthorblockN{Abdul Basit, Maha Nawaz, Muhammad Shafique}
\IEEEauthorblockA{\textit{eBRAIN Lab, Division of Engineering}
\textit{New York University (NYU) Abu Dhabi}, Abu Dhabi, UAE \\
\{abdul.basit, mzn2386, muhammad.shafique\}@nyu.edu }
\vspace{-2em}
}


\maketitle
\input{Sections/abstract}

\input{Sections/Introduction}
\input{Sections/Related_Work}

\input{Sections/Methodology}

\input{Sections/conclusion}

\bibliographystyle{IEEEtran}
\bibliography{cite}

\end{document}

%% file: Sections/abstract.tex
\begin{abstract}

Non-invasive brain-computer interfaces (BCIs) have the potential to enable intuitive control of prosthetic limbs for individuals with upper limb amputations. However, existing EEG-based control systems face challenges related to signal noise, classification accuracy, and real-time adaptability. In this work, we present BRAVE, a hybrid EEG and voice-controlled prosthetic system that integrates ensemble learning-based EEG classification with a human-in-the-loop (HITL) correction framework for enhanced responsiveness. Unlike traditional electromyography (EMG)-based prosthetic control, BRAVE aims to interpret EEG-driven motor intent, enabling movement control without reliance on residual muscle activity. To improve classification robustness, BRAVE combines Long Short-Term Memory (LSTM), Convolutional Neural Network (CNN), and Random Forest models in an ensemble framework, achieving a classification accuracy of 96\% across test subjects. EEG signals are preprocessed using a bandpass filter (0.5–45 Hz), Independent Component Analysis (ICA) for artifact removal, and Common Spatial Pattern (CSP) feature extraction to minimize contamination from electromyographic (EMG) and electrooculographic (EOG) signals. Additionally, BRAVE incorporates automatic speech recognition (ASR) to facilitate intuitive mode switching between different degrees of freedom (DOF) in the prosthetic arm. The system operates in real time, with a response latency of 150 ms, leveraging Lab Streaming Layer (LSL) networking for synchronized data acquisition. The system is evaluated on an in-house fabricated prosthetic arm and on multiple participants highlighting the generalizability across users. The system is optimized for low-power embedded deployment, ensuring practical real-world application beyond high-performance computing environments. Our results indicate that BRAVE offers a promising step towards robust, real-time, non-invasive prosthetic control.

\end{abstract}


%% file: Sections/Introduction.tex
\section{Introduction}

According to the World Health Organization, approximately 2.5 million people worldwide live with upper limb amputations, with a significant portion residing in developing countries, where access to prosthetic care is severely limited \cite{world2022global}. Globally, about 57.7 million people suffer from limb amputations, primarily due to traumatic causes, including falls (36.2\%), road injuries (15.7\%), and mechanical forces (10.4\%) \cite{doi:10.1177/0309364620972258}. Among these, transhumeral amputations account for 2.5\% of all cases \cite{inkellis2018incidence}. The residual limb is often underutilized in daily activities, leading to muscle atrophy and ankylosis, further complicating prosthetic control \cite{fitzgibbons2015functional}. This deterioration significantly impacts surface electromyography (sEMG)-based prosthetic systems, which depend on residual muscle activity to function, thus limiting their effectiveness in high-level amputations. 

This has driven the need for alternative control systems, particularly non-invasive brain-computer interfaces (BCIs), which have emerged as a promising approach for prosthetic limb control \cite{mcfarland2018brain}. Electroencephalography (EEG)-based BCIs provide a direct neural interface, bypassing reliance on residual musculature, making them particularly beneficial for high-level amputees. However, several challenges hinder the widespread adoption of EEG-based prosthetic control, including low signal-to-noise ratio (SNR), artifacts from eye blinks and muscle movements, high inter-subject variability, and latency in real-time applications \cite{he2021online}. Furthermore, in low-income countries, only 5\%–15\% of individuals have access to prosthetic care, making affordability a critical issue \cite{Abbady2022}. 3D printing technologies offer a cost-effective solution to manufacture personalized prosthetic limbs, potentially making EEG-driven prosthetic systems more accessible in resource-constrained environments.

\textit{Challenges in EEG-Based Prosthetic Control}: Non-invasive EEG signals suffer from {low spatial resolution and susceptibility to contamination} from non-neural sources, such as {electromyographic (EMG) activity from facial and jaw muscles} and {electrooculographic (EOG) artifacts due to eye movements} \cite{jiang2014intuitive, krusienski2008toward}. These confounds often {skew classification models}, leading to {overestimated accuracy} in systems that do not explicitly account for them \cite{fatourechi2007emg}. Furthermore, EEG-based BCIs rely on {motor imagery (MI) paradigms}, which require {extensive user training} and are highly {subject-dependent} \cite{lotte2018flaws}. Most current systems fail to {generalize across users}, necessitating time-consuming calibration for each individual \cite{schirrmeister2017deep, craik2019deep}.  

Additionally, {real-time control remains a major bottleneck}. Many EEG classification pipelines require {computationally expensive preprocessing}, causing {delays that hinder usability in dynamic prosthetic applications} \cite{lawhern2018eegnet, lopez2018continuous}. While recent advancements in {ensemble learning} and {deep neural networks} have improved classification accuracy, {their deployment on low-power, embedded systems remains a challenge} \cite{Millan2010}.

The primary challenge addressed in this paper is the development of a robust, non-invasive brain-computer interface (BCI) for controlling prosthetic arms using EEG signals. This is crucial for creating advanced prosthetics that mimic natural limb functions and improve the quality of life for individuals with limb loss or motor impairments \cite{Capsi-Morales2023-ir}. While progress in prosthetic technology has been made, achieving seamless, intuitive control remains difficult due to the inherent noise and complexity of EEG signals, complicating real-time interpretation. Prior research \cite{9328561} has shown the potential of BCIs for controlling external devices via neural activity, but current solutions often face limitations in accuracy, latency, and usability.



\begin{table*}[b]
    \centering
    \caption{Comparison of Our Approach with Existing EEG-Based Prosthetic Control Systems}
    \label{tab:comparison}
    \begin{tabular}{|l|c|c|c|c|c|}
        \hline
        \textbf{System} & \textbf{Control Signals} & \textbf{ML Model} & \textbf{Artifact Removal} & \textbf{Latency (ms)} & \textbf{Generalization} \\
        \hline
        NeuroLimb \cite{NeuroLimb} & EEG + EMG & SVM & Not Reported & 250 & Limited \\
        MindArm \cite{Nawaz_2024} & EEG & CNN & Not Reported & 200 & Limited \\
        Hybrid EEG-EOG \cite{Hybrid_EEG_EOG} & EEG + EOG & CSP + LDA & ICA & 230 & Limited \\
        Ensemble BCI \cite{EnsembleBCI} & EEG & LSTM + RF & Not Reported & 180 & Moderate \\
        \textbf{BRAVE (Ours)} & EEG + Voice & LSTM + CNN + RF & ICA + CSP & \textbf{150} & \textbf{Moderate-High} \\
        \hline
    \end{tabular}
\end{table*}

To address the limitations of current EEG-based prosthetic control systems, this paper presents {BRAVE}, a {hybrid EEG and voice-controlled prosthetic system} that integrates {ensemble learning-based EEG classification} with {a human-in-the-loop control framework} for enhanced responsiveness with the following \textit{novel contributions}:  

\begin{enumerate}
    \item \textbf{EEG-Based Prosthetic Control with Artifact Rejection}: We employ {Independent Component Analysis (ICA)} and {Common Spatial Pattern (CSP) feature extraction} to {minimize contamination from EMG and EOG artifacts}, ensuring that classification models rely on {pure EEG signals} rather than unintended muscle movements.  

    \item \textbf{Ensemble Learning for Robust Classification}: We combine {Long Short-Term Memory (LSTM), Convolutional Neural Network (CNN), and Random Forest classifiers} to enhance {classification robustness}, achieving a {96\% accuracy rate} across test subjects.  

    \item \textbf{Real-Time, Low-Latency Prosthetic Control}: The system is optimized for {low-latency real-time control (150 ms response time)} using {Lab Streaming Layer (LSL) networking} to synchronize EEG signal acquisition and classification.  

    \item \textbf{Hybrid EEG-Voice Control for Intuitive Multi-DOF Movement}: {Automatic Speech Recognition (ASR)} allows for {mode-switching between different DOFs}, improving {usability in real-world scenarios} where EEG alone may be insufficient.  

    \item \textbf{Generalizability and Embedded System Deployment}: The system is tested on {multiple participants}, highlighting {inter-subject generalization} and is optimized for {low-power embedded deployment}, ensuring feasibility beyond high-performance computing environments.  
\end{enumerate}

The remainder of this paper is structured as follows: {Section II} reviews related work in {EEG-based prosthetic control, hybrid BCI systems, and deep learning approaches for neural classification}. {Section III} describes the {proposed methodology}, detailing the {EEG acquisition, preprocessing pipeline, machine learning framework, and prosthetic hardware integration}. {Section IV} presents the {experimental setup and evaluation metrics}, followed by {results and discussion in Section V}. Finally, {Section VI} concludes with future directions for {improving inter-subject EEG adaptation and embedded deployment optimization}.

These contributions overcome key limitations in existing systems by enhancing signal accuracy, reducing latency, and improving usability, leading to more reliable prosthetic control. Our experiments show a significant improvement, with the integration of the OpenBCI platform and ensemble learning models achieving 96\% classification accuracy across 4 test subjects using a generalized model.

%% file: Sections/Related_Work.tex
\section{Related Work}

\subsection{Hybrid EEG-EMG Systems for Prosthetic Control}

Recent advancements in {brain-controlled prosthetics} have led to the development of {hybrid control systems} that integrate EEG and EMG signals to enhance the precision and responsiveness of prosthetic devices. Notable examples include \textit{NeuroLimb} \cite{NeuroLimb} and other hybrid systems \cite{Hybrid_EEG_EMG, EEG-EMG}, which demonstrate the benefits of combining neural and muscular signals to improve gesture recognition and motor intent classification for transhumeral amputees. Studies show that hybrid EEG-EMG models significantly outperform single-modal approaches in tasks requiring fine motor control \cite{HybridReview, EMGEEGComparison}.  

The Hybrid EEG-EMG framework has been particularly beneficial for above-elbow amputees, where sEMG signals alone are insufficient for fine motor control \cite{Hybrid_EEG_EMG}. By utilizing EEG for higher-level intent recognition and EMG for low-level muscle activation, such approaches have improved classification accuracy and robustness \cite{HybridProstheticControl}. Moreover, \textit{MindArm} \cite{Nawaz_2024} employs EEG-driven deep learning models for real-time, cost-effective prosthetic control, making BCIs more accessible for real-world applications.

Beyond EEG-EMG systems, hybrid EEG-EOG (Electrooculography) interfaces have been explored for prosthetic and assistive device control. The {Hybrid EEG-EOG system} \cite{Hybrid_EEG_EOG} leverages Common Spatial Pattern (CSP) algorithms to fuse brain and eye movement signals, enabling more efficient prosthetic command execution, particularly for individuals with limited voluntary muscle control.

\begin{figure*}[!t]
    \centering
    \includegraphics[width=1\linewidth]{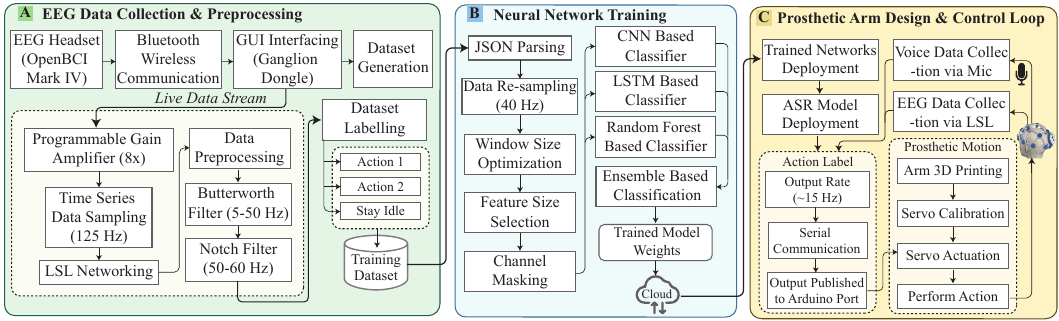}
    \caption{BRAVE System Overview: The methodology consists of three stages: (A) EEG Data Collection \& Preprocessing, where EEG signals are captured using the OpenBCI Mark IV headset, then amplified, filtered, and streamed for analysis and dataset generation; (B) Neural Network Training, where EEG data is classified using an ensemble of DL models, with the window size optimized and models stored in the cloud for deployment; and (C) Prosthetic Arm Design \& Control, where the models are deployed on hardware to classify live EEG data, generating control commands for prosthetic movement. The 3D-printed prosthetic arm, tailored to user needs, operates with 3 DOF via an Arduino-controlled servo system.}
    \label{fig:methodology}    
\end{figure*}

\subsection{Machine Learning in EEG-Based Prosthetic Control}
The integration of machine learning (ML) techniques has significantly improved real-time classification of EEG signals, reducing response latency and enhancing user adaptability. Traditional classifiers such as {Support Vector Machines (SVM)} and {Random Forest (RF)} have demonstrated effectiveness in EEG classification \cite{ML_EEG_SVM_RF}, but recent advances favor deep learning approaches like {Long Short-Term Memory (LSTM)} and {Convolutional Neural Networks (CNN)} \cite{DeepBCI_Review, EnsembleBCI}. These models capture temporal dependencies in EEG time-series data, leading to higher classification accuracy for brain-controlled prosthetic systems \cite{EEG_CNN_LSTM}. Additionally, non-invasive BCIs integrated with {Gradient Boosting algorithms} have exhibited improvements in autonomous robotic arm control \cite{app122110813}, supporting adaptive learning strategies that enhance user-specific performance over time.  

\subsection{Real-Time BCI Control for Assistive Robotics}

aBCIs have extended beyond prosthetic control, enabling real-time interaction with assistive robots. For example, EEG-based systems have been used to control {robotic arms and quadruped robots}, including {Boston Dynamics' Spot} \cite{s24010080}. Additionally, shared control architectures, which balance user input with system autonomy, further improve usability in complex environments \cite{9636261}. Multi-modal BCIs, integrating EEG, EOG, and ASR, have been explored for seamless, intuitive prosthetic control \cite{HybridASR_BCI}.

\subsection{Comparison with Existing Systems}

A comparison of our approach with recent EEG-based and hybrid prosthetic control systems is presented in Table~\ref{tab:comparison}.

\noindent Our BRAVE system differs from existing methods in the following ways:
\begin{itemize}
    \item It incorporates {ICA + CSP} for artifact rejection, reducing contamination from {EMG and EOG signals}.
    \item It achievs a {lower response latency (150 ms)}, making it more {suitable for real-time prosthetic applications}.
    \item It optimizes for {real-world generalization} by including {multi-subject evaluation}.
\end{itemize}

\noindent These contributions position BRAVE as a state-of-the-art non-invasive EEG prosthetic control system, demonstrating high usability, robustness, and adaptability for real-world applications.

%% file: Sections/Methodology.tex
\section{Methodology} 
The BRAVE system ({Brain-Controlled Prosthetic Arm with Voice Integration and Embodied Learning for Enhanced Mobility}) integrates {EEG-based control}, {automatic speech recognition (ASR)}, and {machine learning-driven classification} to provide an intuitive, real-time prosthetic control framework. The methodology consists of four primary components: {EEG signal acquisition and preprocessing}, {ensemble learning-based classification}, {voice command integration}, and {prosthetic arm actuation}. This section provides a detailed breakdown of each component.


In this section, we detail the proposed BRAVE system, including the workflow for EEG dataset generation, refinement, and training. Additionally, we discuss the design and control strategy of the prosthetic arm, which enables the execution of various tasks. The Figure \ref{fig:methodology} showcases the system methodology.

\subsection{EEG Data Collection \& Preprocessing}
The EEG data collection process in our system is critical for ensuring the accuracy and effectiveness of the prosthetic arm control. We utilize the Ultracortex Mark IV headset \cite{MarkIV} along with the 16-channel Cyton + Daisy Biosensing boards \cite{Cyton_board}, chosen for their balance of affordability, practicality, and enhanced capability to stream 16 channels of EEG data. This setup provides a significant improvement over the 4-channel Ganglion board \cite{GANGLION_board}, allowing for more comprehensive data collection and finer resolution in capturing neural activity. Unlike other EEG devices such as the Emotiv EPOC X  \cite{Emotiv}, which require saline or gel-based electrodes, the Ultracortex Mark IV headset operates with dry comb electrodes. This design choice significantly enhances the practicality and ease of use in daily applications, making it a suitable option for long-term use without the discomfort and maintenance associated with wet electrodes.

The Cyton + Daisy boards extend the number of EEG channels, enabling detailed brain activity capture across multiple scalp regions. Electrodes are positioned using the 10-20 system, covering areas like Fp1, Fp2, F3, F4, C3, C4, etc. to optimize detection of brainwave frequencies (alpha, beta, theta) across the frontal, central, and occipital regions. 
\begin{figure}[ht]
    \centering
    \includegraphics[width=1\linewidth]{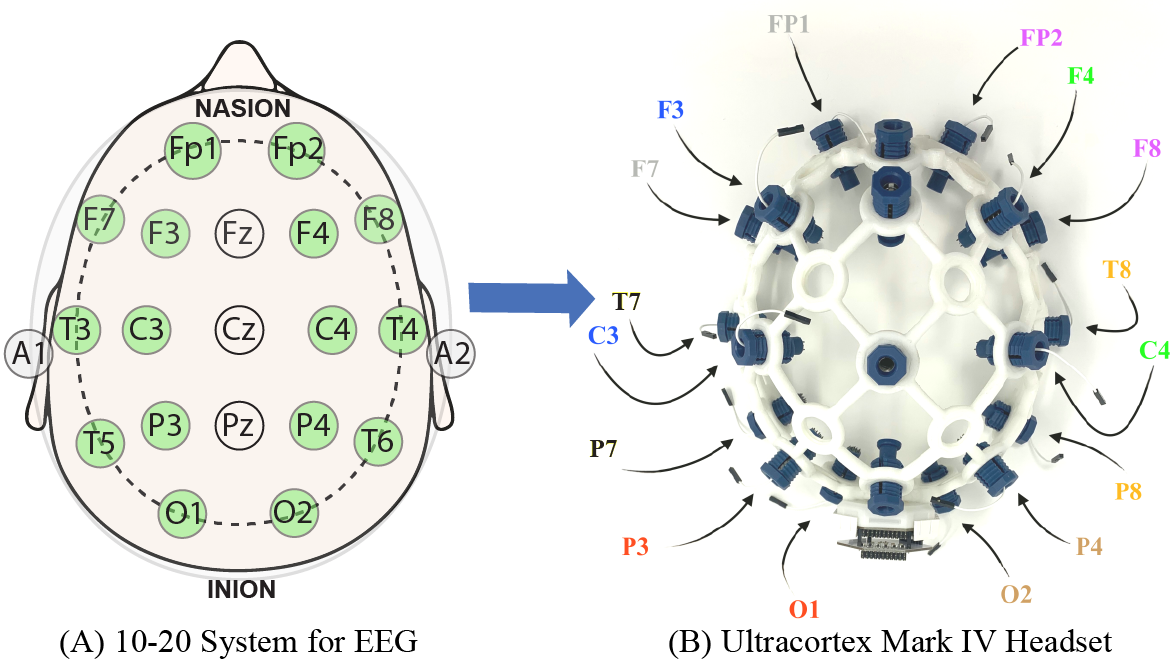}
    \caption{EEG Data Collection Process: (A) The 10-20 system for EEG electrode placement, highlighting standard locations for brainwave detection. (B) The Ultracortex Mark IV headset equipped with 16 channels for comprehensive EEG acquisition.}
    \label{fig:datacollection}    
\end{figure}

\begin{figure}[ht]
  \centering
    \includegraphics[width=1\linewidth]{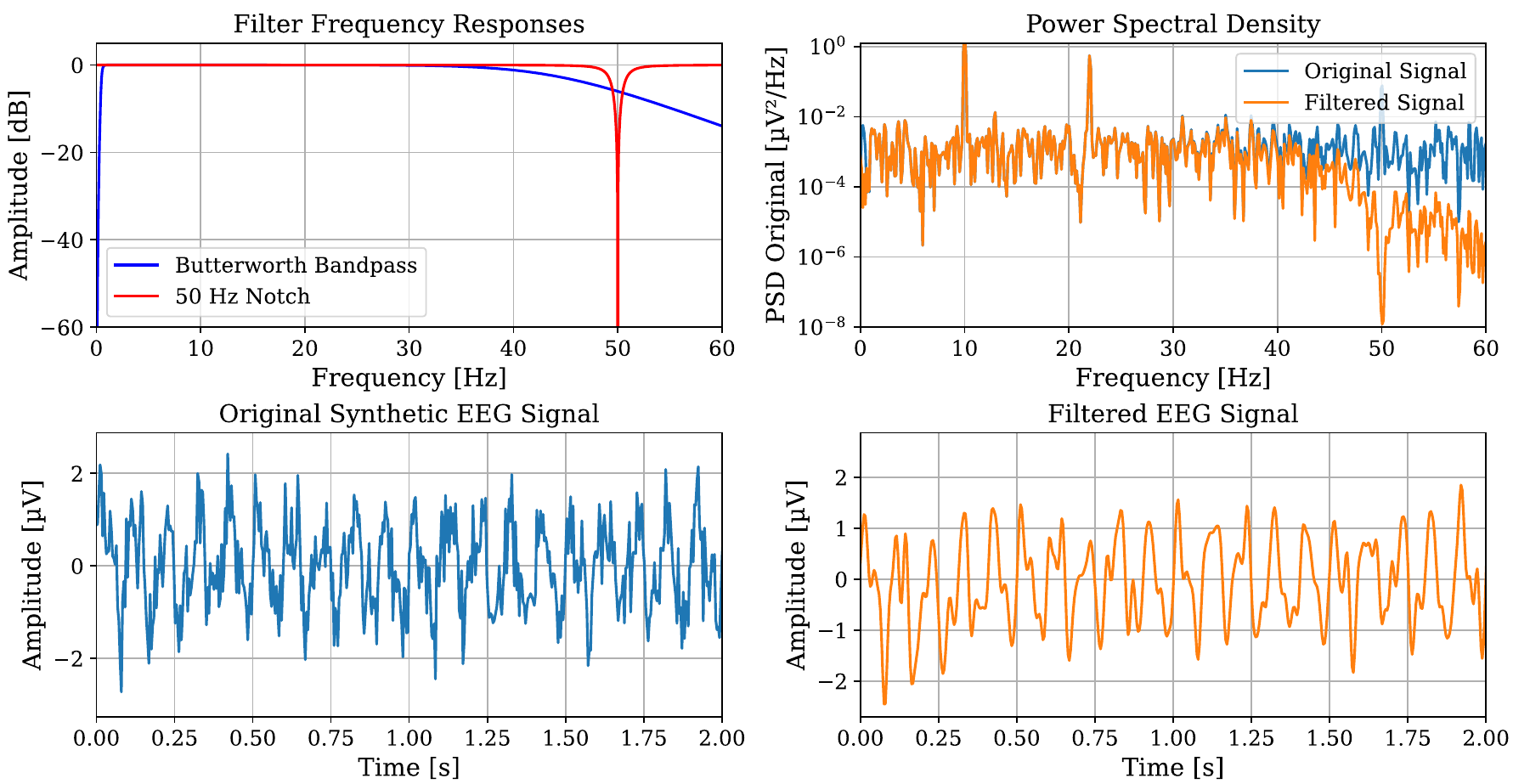} 
    \caption{(a) Frequency responses of the 6th order Butterworth bandpass filter (0.5–45Hz) and the 50Hz notch filter, showing their attenuation profiles.
(b) Power Spectral Density (PSD) of the original and filtered signals, highlighting the reduction of noise and elimination of the 50Hz interference after filtering.
(c) Time-domain plot of the original EEG signal with alpha and beta waves, noise, and 50Hz interference.
(d) Time-domain plot of the filtered EEG signal after applying both filters, demonstrating improved signal clarity and preservation of key EEG components.}
    \label{fig:filter}    

  \end{figure}

This comprehensive coverage allows for more accurate and reliable interpretation of EEG signals, which is crucial for precise prosthetic arm control. Figure \ref{fig:datacollection} illustrates the data collection pipeline. We preprocess the EEG data by amplifying it and then applying filters: a Butterworth filter to retain relevant frequency bands and a notch filter to eliminate powerline interference. This reduces artifacts and enhances the signal-to-noise ratio, which is crucial for accurate classification by the neural network. The filtering process is illustrated in Figure \ref{fig:filter}.

We utilize the OpenBCI GUI software which offers extensive control over filter settings, channel selection, fine-tuning PGA Gain, bias, and other hardware settings for each channel along with real-time data visualization. The Programmable Gain Amplifier (PGA) is essential in EEG systems for amplifying the weak brain signals, typically ranging from 1 to 100 µV, to levels suitable for processing. The PGA allows dynamic adjustment of the gain to optimize the signal-to-noise ratio (SNR), ensuring clear and accurate signal capture. The output voltage $V_{out}$ of the PGA is given by: $V_{out} = A * V_{in}$; where is the input EEG signal, and $A$ is the programmable gain set to 8x after experimentation.

\begin{figure}[h]
\begin{minipage}{0.5\linewidth}
\centering
\includegraphics[width=\linewidth]{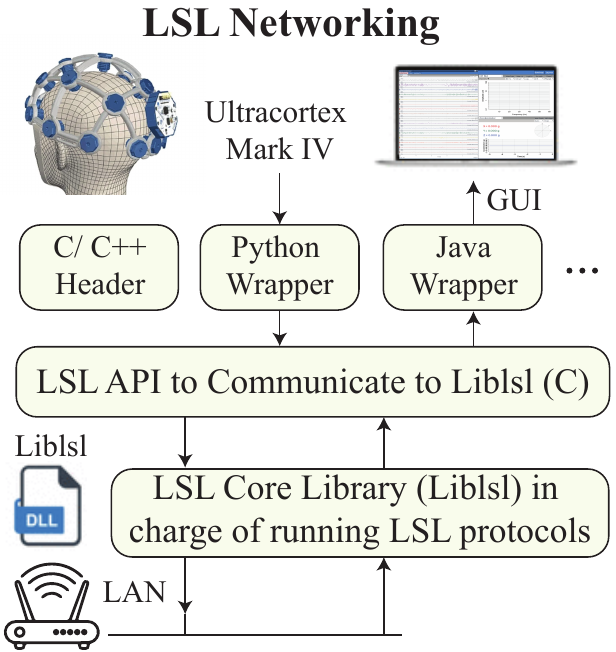}
\end{minipage}%
\hfill
\begin{minipage}{0.48\linewidth}
\caption{Using Lab Streaming Layer (LSL) Networking \cite{Kothe2024} instead of UDP significantly enhances system performance for EEG-based prosthetic control. LSL facilitates real-time streaming and synchronization of EEG data across devices, providing a consistent sample rate around 125 Hz—crucial for capturing rapid EEG dynamics necessary for accurate neural pattern detection and precise prosthetic control.}

\end{minipage}
\end{figure}
 In contrast, UDP's lower sample rate of around 25 Hz is insufficient for capturing higher-frequency EEG components vital for motor control. LSL ensures minimal latency and accurate time-stamping, enabling synchronized data streams in complex setups involving multiple data sources. This results in improved signal fidelity, EEG classification accuracy, and overall system responsiveness, making LSL ideal for real-time applications like BCIs.

Integrating these tools ensures a seamless interface for data collection and analysis, enhancing the robustness and reliability of our EEG-based prosthetic control system. By leveraging the capabilities of the Ultracortex Mark IV and Cyton + Daisy boards, we achieve higher-resolution brainwave data, facilitating more nuanced and responsive control of the prosthetic arm.

\subsection{Independent Component Analysis (ICA) for Artifact Removal}

To improve EEG signal quality and eliminate non-neural artifacts, we applied \textbf{Independent Component Analysis (ICA)}, a commonly used blind source separation technique for EEG preprocessing \cite{jung2000removing}. The ICA decomposition was performed using the \texttt{MNE-Python} library, leveraging the FastICA algorithm to extract statistically independent components from the EEG recordings. 

\subsubsection{Preprocessing and ICA Application}
The EEG signal was first preprocessed as follows:
\begin{itemize}
    \item \textbf{Bandpass filtering (1–40 Hz):} Eliminates low-frequency drift and high-frequency noise.
    \item \textbf{Notch filtering (50 Hz):} Removes power-line interference.
    \item \textbf{Montage assignment:} A synthetic montage with jitter correction was applied to prevent topographic errors.
    \item \textbf{Epoching:} Data was segmented into 1-second epochs to facilitate spatial pattern analysis.
\end{itemize}

\noindent The ICA decomposition was applied to the filtered EEG signals with the number of components set to match the number of available channels. The resulting independent components were visually inspected to identify and remove artifacts associated with eye blinks, muscle activity, and line noise \cite{pion2018icabased}. The identified artifacts were then excluded, and the cleaned EEG signal was reconstructed.

\subsubsection{Component Visualization}
\begin{figure}[h]
    \centering
    \includegraphics[width=0.5\textwidth]{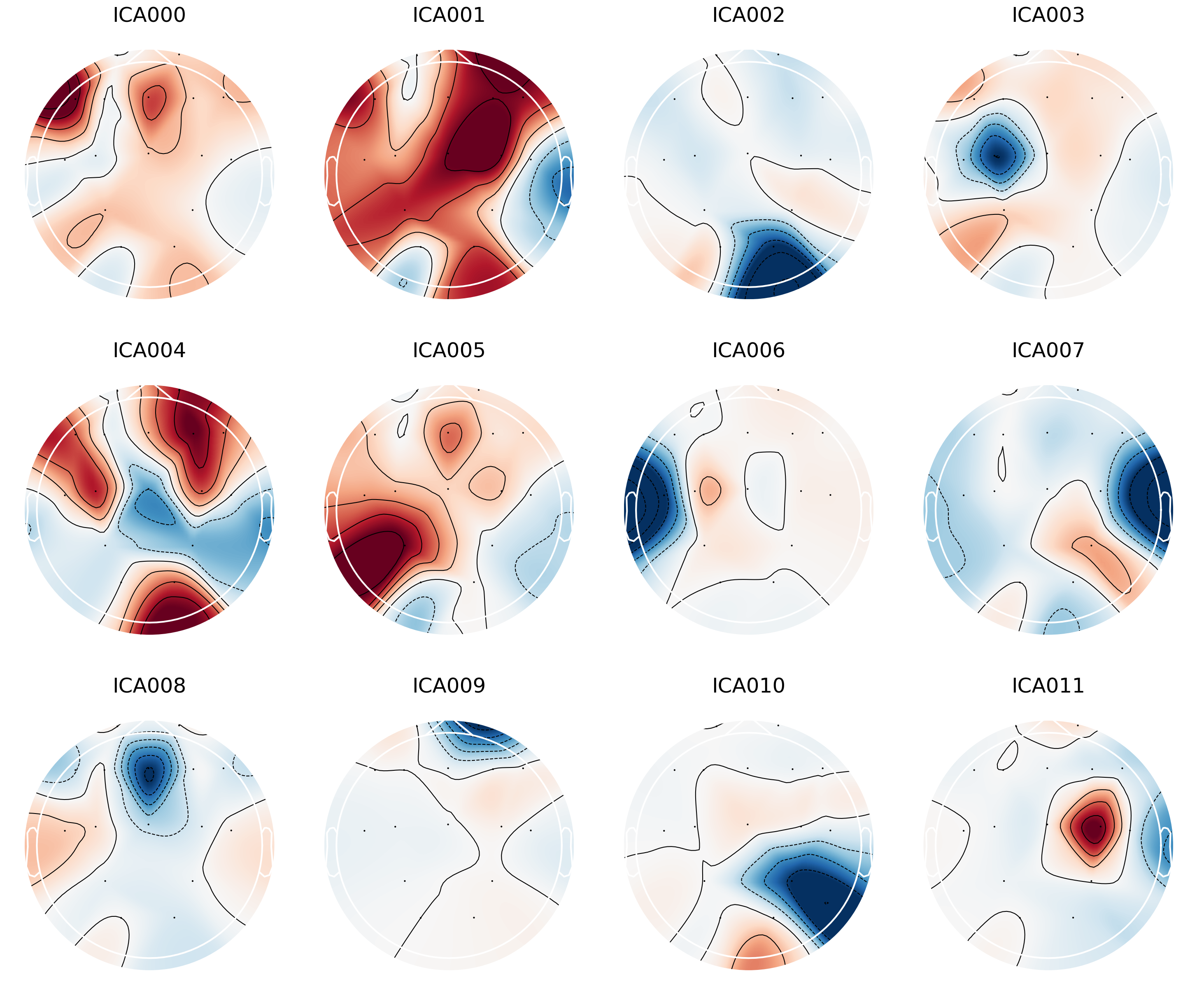}
    \caption{ICA Component Topographies (IC 000 – IC 011): Scalp maps of the 12 independent components extracted from the 16-channel, 1–40 Hz-filtered EEG. Red (positive weights) and blue (negative weights) indicate each component’s contribution across electrodes; darker colours represent larger absolute loadings. Components with strong, focal frontal activity (e.g. IC 001) are characteristic of ocular artefacts, whereas more distributed bilateral patterns (e.g. IC 004, IC 007) are likely of neural origin. These maps guide manual rejection by highlighting artefact-dominated ICs.}
    \label{fig:ica_components}
\end{figure}

The extracted ICA components are illustrated in Figure~\ref{fig:ica_components}. These topographic maps highlight spatial distributions of the independent components, where red and blue regions indicate sources of positive and negative activations, respectively. Components exhibiting characteristic eye-blink or muscle artifact patterns were manually rejected.

\noindent The cleaned EEG signals were subsequently used for feature extraction and classification, ensuring that motor-related EEG activity remained the primary source of control information.

\subsection{Common Spatial Pattern (CSP) for Feature Extraction}

To enhance the discrimination of EEG motor imagery signals, we employed \textbf{Common Spatial Pattern (CSP)} filtering, a widely used method for optimizing the variance separation between two distinct classes in Brain-Computer Interface (BCI) applications \cite{ramoser2000optimal}. CSP is particularly effective for motor imagery tasks, as it extracts spatial filters that maximize variance for one class while minimizing it for another, thereby improving classification performance \cite{lotte2018review}.

\subsubsection{CSP Computation}

The CSP algorithm was applied to EEG data following ICA-based artifact removal. The key steps involved in CSP computation include:
\begin{itemize}
    \item \textbf{Epoching:} EEG signals were segmented into 1-second epochs to ensure consistency in feature extraction.
    \item \textbf{Covariance Estimation:} Covariance matrices were computed for each epoch, representing spatial correlations across channels.
    \item \textbf{Eigenvalue Decomposition:} A transformation matrix was derived to maximize the variance ratio between two conditions (e.g., left-hand vs. right-hand motor imagery).
    \item \textbf{Feature Projection:} The EEG signals were projected onto the CSP filters, yielding discriminative features for classification.
\end{itemize}

\subsubsection{CSP Pattern Visualization}

\begin{figure}[h]
    \centering
    \includegraphics[width=0.48\textwidth]{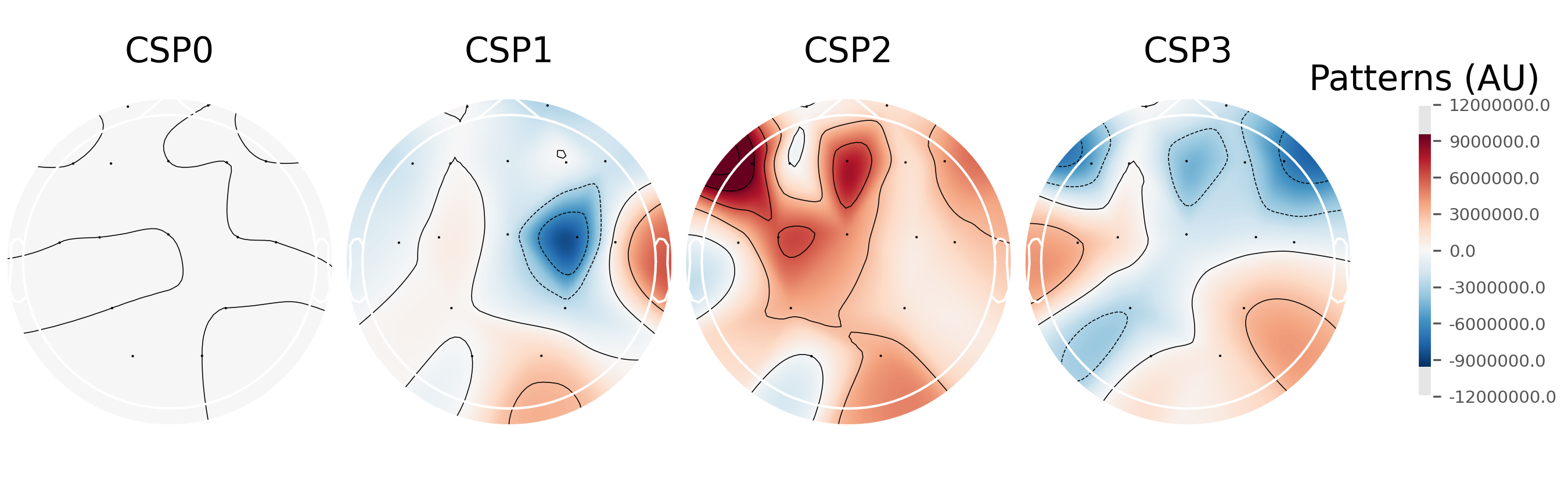}
    \caption{CSP Spatial Patterns (CSP 0 – CSP 3): Topographic distribution of the four most discriminative common spatial patterns computed between the label classes. CSP 0 captures negligible discriminative power (near-uniform map), whereas CSP 1–CSP 3 emphasise centro-parietal regions with reciprocal polarity, reflecting sensor pairs whose variance ratios differ most across classes. Red areas contribute positively and blue areas negatively to the corresponding CSP feature, scaled in arbitrary units (AU).}
    \label{fig:csp_patterns}
\end{figure}

Figure~\ref{fig:csp_patterns} illustrates the CSP spatial patterns corresponding to the most significant components extracted from the EEG dataset. The red and blue regions indicate areas of increased and decreased signal variance, respectively, highlighting the discriminative spatial distribution of motor-related activity.

\noindent The extracted CSP features were subsequently fed into the machine learning classification pipeline, enhancing the system's ability to distinguish between different motor intention states.

\subsection{Neural Network Training and Ensemble Learning}
\begin{figure}[H]
    \centering
    \includegraphics[width=1\linewidth]{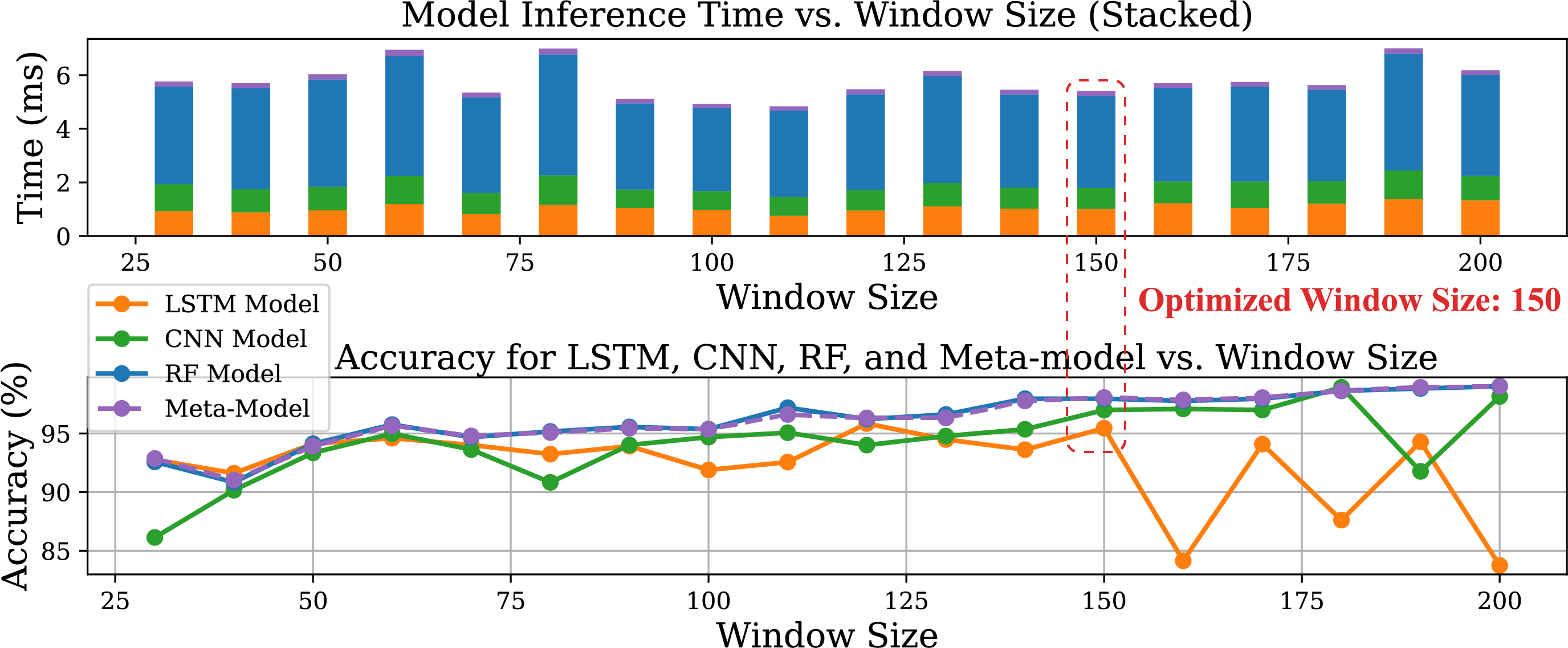}
    \caption{The figure illustrates stacked inference time for each model (LSTM, CNN, RF, and Meta-model) across various window sizes, along with the corresponding test accuracies. This comparison highlights the trade-off between performance and inference time as the window size increases.}
    \label{fig:window}    
\end{figure}

We employ an ensemble of models—including LSTM, CNN, and Random Forest-based classifiers—for accurate and robust EEG signal interpretation in real-time prosthetic limb control. This approach enhances system performance by combining the strengths of each model, resulting in improved classification accuracy and robustness. Our analysis shows that a window size of 150 offers an optimal balance between accuracy and inference time. While CNN and RF models remain consistent, LSTM accuracy fluctuates beyond this size, peaking at 150. This window size ensures high accuracy across models with manageable inference time, ideal for real-time EEG applications as illustrated in Figure \ref{fig:window}.

\subsubsection{LSTM-Based Classification Model}
The Long Short-Term Memory (LSTM) network is utilized to capture temporal dependencies in EEG time-series data, which is crucial for recognizing brainwave patterns associated with specific motor commands. Our LSTM architecture consists of 64 hidden layers designed to retain information from previous time steps, enabling the network to learn complex temporal dynamics. Dropout layers are incorporated to mitigate overfitting, and model outputs the classification results. The LSTM model is trained on filtered EEG data, leveraging its ability to model sequential data effectively.
\subsubsection{CNN-Based Classification Model}
To complement the temporal modeling of the LSTM, we utilize a Convolutional Neural Network (CNN) to capture spatial patterns within the EEG data. The CNN is adept at detecting localized features across multiple EEG channels, offering a different perspective on the signal's structure. By treating windowed EEG data as multi-channel images, the CNN applies convolutional layers to extract spatial features, followed by pooling layers to reduce dimensionality and enhance feature robustness. The network concludes with fully connected layers that output the classification results. The CNN model is trained in parallel with the LSTM, and its outputs contribute to the ensemble learning framework.

\subsubsection{Random Forest-Based Classification Model}

We incorporate a Random Forest classifier into our ensemble to leverage its capability to handle high-dimensional data and model complex, non-linear relationships inherent in EEG signals. The Random Forest enhances classification performance by constructing an ensemble of decision trees, each trained on random subsets of the data and features, which improves robustness and reduces overfitting. We extract statistical features from the preprocessed EEG signals—such as mean, variance, standard deviation, maximum, and minimum values—that serve as inputs to the Random Forest classifier. This model provides a complementary perspective to the deep learning models, capturing patterns that may be missed by neural networks.

\begin{figure}[ht]
    \centering
    \includegraphics[width=1\linewidth]{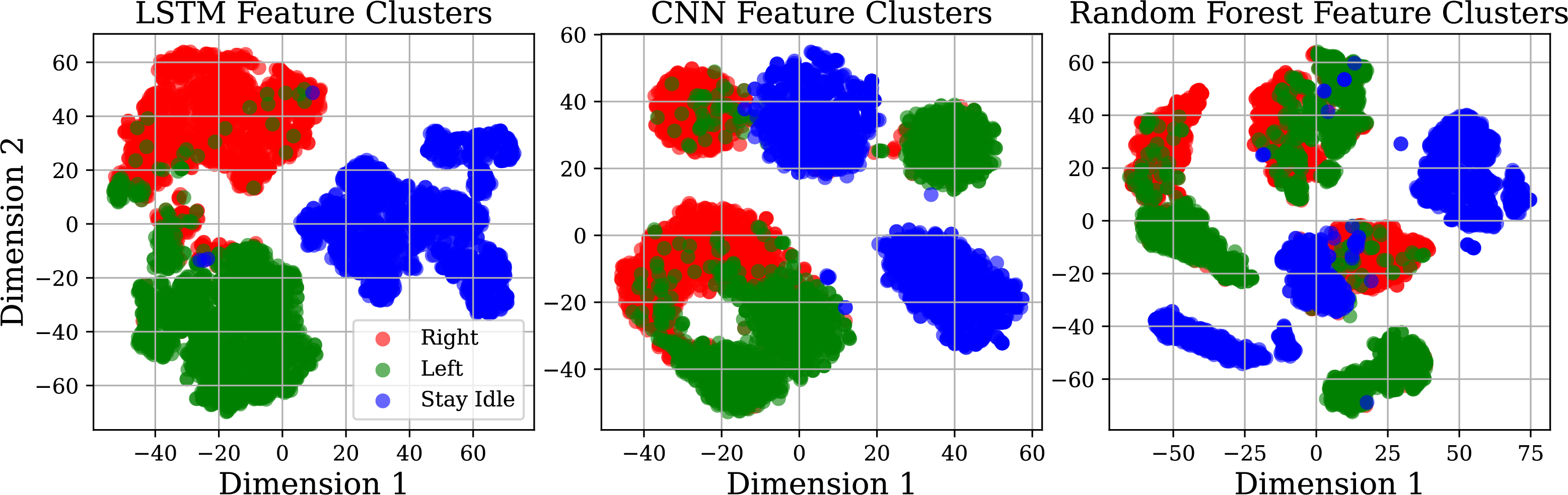}
    \caption{t-SNE Feature Clustering for classifiers illustrating the two-dimensional embedding of high-dimensional features. The distinctness of clusters indicates each model's ability to separate classes, with LSTM capturing temporal dependencies, CNN capturing spatial patterns, and Random Forest modeling complex non-linear relationships.}
    \label{fig:SNE}    
\end{figure}

\subsubsection{Ensemble Learning}

The ensemble learning approach combines LSTM, CNN, and Random Forest classifiers to create a more robust EEG signal classification system. Each model contributes its strengths—temporal patterns (LSTM), spatial features (CNN), and non-linear relationships (Random Forest) shown in Figure \ref{fig:SNE}. A logistic regression meta-classifier optimally integrates these outputs, enhancing accuracy and generalization while reducing overfitting.

\begin{figure}[ht]
    \centering
    \includegraphics[width=1\linewidth]{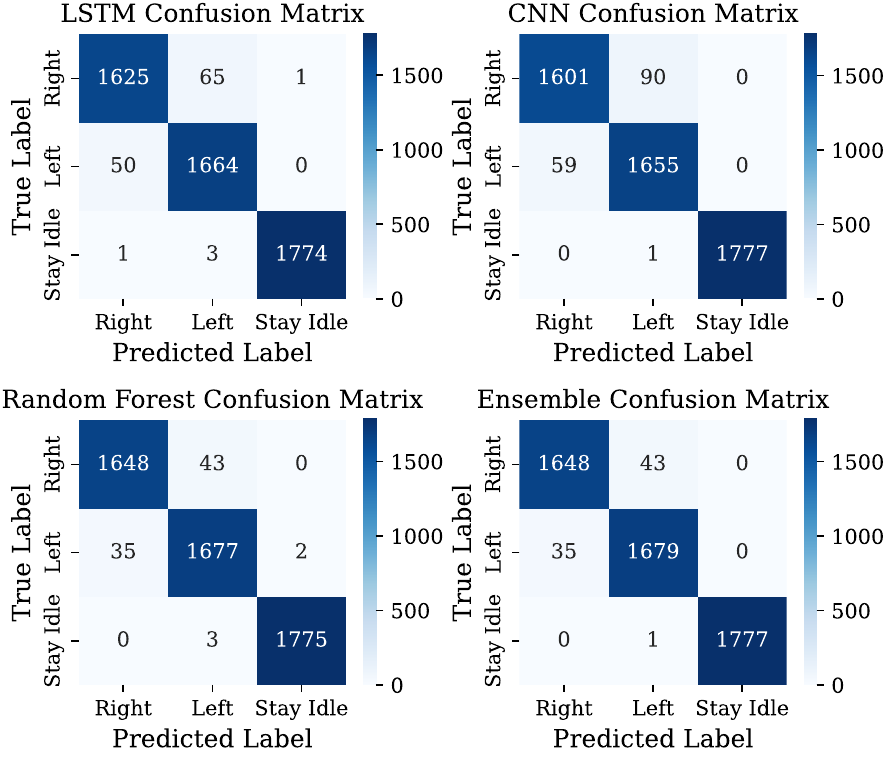}
    \caption{Confusion matrices for each individual classifier and the ensemble model illustrate classification performance. The ensemble model demonstrates the best performance, with the highest accuracy and lowest misclassification rates, highlighting its ability to combine the strengths of all models for robust EEG signal classification on test dataset.}
    \label{fig:confusion}    
\end{figure}

The ensemble is tested extensively, both offline (using a validation dataset) and online (in real-time scenarios), to ensure that it meets the performance criteria for accuracy, responsiveness, and reliability. The final ensemble model is saved and deployed within the prosthetic system, where it continuously processes incoming EEG data to control the prosthetic limb in real time. Figure \ref{fig:confusion} shows the test results.

\begin{figure*}[ht]
    \centering
    \includegraphics[width=0.9\linewidth]{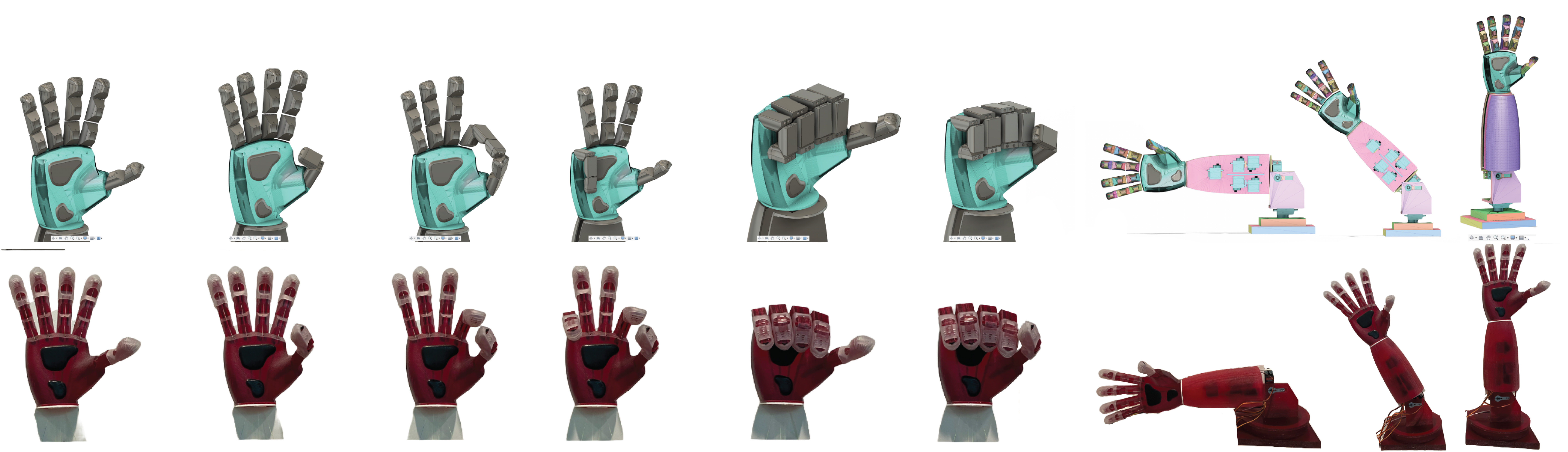}
    \caption{The design and physical prototype of the prosthetic hand. The top row (A) showcases various orientations of the 3D modeled hand, illustrating different possible gestures and positions. The bottom row (B) displays the corresponding physical prototypes, which were 3D printed using the Stratasys J750 printer. The servo motors positioned for controlling the fingers, elbow, and base provide 3 DOF. Vero and Tango materials were used for 3D printing. }
    \label{fig:handpositions}    
\end{figure*}

\subsection{Prosthetic Arm Design and Manufacturing}

The prosthetic arm was designed and manufactured with a focus on optimizing functionality, durability, aesthetics, and cost. Utilizing Autodesk Fusion 360 for precise 3D modeling and simulation, we engineered the arm to control three main degrees of freedom (DOF): flexion/extension of the elbow and independent finger movements for gripping and pinching. The design underwent iterative refinement to enhance mechanical efficiency and user interface.

\subsubsection{3D Printing and Prototyping}

After finalizing the design, the components were 3D printed using the Stratasys J750 printer, renowned for its precision and multi-material capabilities. The GrabCAD interface streamlined the printing process, ensuring each part had accurate material properties and dimensions. The assembled prosthetic arm comprises the hand, forearm, and base, with the hand featuring individually segmented fingers controlled by five embedded servo motors, enabling precise movements such as gripping and pinching as illustrated in Figure \ref{fig:handpositions}.

\subsection{Prosthetic Arm Control}

The prosthetic arm control system integrates trained neural network models with an Arduino microcontroller to drive servos for precise movement. This subsection describes how control commands are generated, transmitted, and executed.

\subsubsection{Model Deployment}

We deployed the trained ensemble learning models on a high-performance laptop capable of efficiently handling real-time EEG signal processing and classification. For prototyping, a Lambda Tensorbook with an NVIDIA 3080 Ti GPU was used to ensure minimal latency.

\subsubsection{Generating Action Labels}
After EEG data collection, the trained models process the incoming data to classify the intended action (e.g., move left, move right, or stay idle). The action label is generated at regular intervals, typically at a rate of 15 Hz, ensuring a responsive system that can adapt to the user’s mental commands in real-time.

\subsubsection{Communication with Arduino}
The next step involves sending the generated action labels to the Arduino microcontroller via a serial port. The communication between the GPU-based laptop and the Arduino is established using a serial communication protocol, which allows for the reliable transmission of control commands. The Arduino is responsible for interpreting these commands and converting them into precise motor signals.

The serial communication process involves the following:
\begin{itemize}  [leftmargin=0pt]
    \item The action label, determined by the neural network models, is sent as a serial data packet to the Arduino.
    \item Upon receiving the data, the Arduino interprets the action label and determines the corresponding movement pattern for the prosthetic arm.
    \item The Arduino sends control signals to the servo motors to execute the desired movement.
\end{itemize}

\subsubsection{Servo Calibration and Arm Movement}
Before deployment, the prosthetic arm's servo motors undergo an initial calibration to ensure alignment to their neutral positions, enabling consistent and accurate movements. This calibration is crucial for addressing any mechanical variations in the assembly. The calibration is performed using CCPM 3 channel servo tester. 

\subsubsection{Automatic Speech Recognition Model Deployment}

\begin{figure}[ht]
    \centering
    \includegraphics[width=0.95\linewidth]{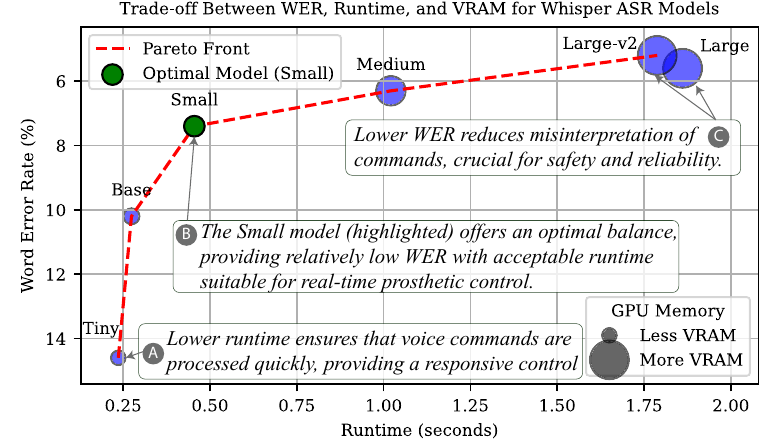}
    \caption{Trade-off Between WER, Runtime, and GPU Memory Consumption for Whisper ASR Models for different Whisper ASR models.}
    \label{fig:whisper}    
\end{figure}

We implemented an Automatic Speech Recognition (ASR) system using the Whisper model to enhance control of the prosthetic arm. Voice commands like "elbow," "arm," and "fingers" switch control modes, while specific hand poses are executed by saying the pose name. While precise movements rely on EEG-based classification, ASR provides intuitive control over the prosthetic's 3 DOF. The Word Error Rate (WER) of the ASR system affects command accuracy, making model selection crucial for reliable control. To address latency issues, the ASR operates in a separate thread, allowing the ensemble model to run concurrently. A Voice Activity Detection (VAD) algorithm ensures the Whisper model is triggered only by speech prompts, optimizing resource usage. Figure \ref{fig:whisper} illustrates the selection process.

\subsection{Human-in-the-Loop Control System for Prosthetic Arm}

The following equations describe the human-in-the-loop control system for a prosthetic arm driven by EEG signals and voice commands.

\subsubsection{EEG Signal Classification}

The ensemble classifier processes the EEG input to predict a control action $\hat{a}(t)$: $\hat{a}(t) = f_{control}(S_{EEG}(t), W)$ where $S_{EEG}(t)$ is the time-varying EEG signal, and $W$ represents the trained weights of the ensemble learning model. The output $\hat{a}(t)$ is the predicted action, such as moving arm in one direction.

\subsubsection{Voice Command Switching}
Voice commands change the degrees of freedom (DOF) of the prosthetic arm: $M(t) = C_{voice}(t)$ where $M(t)$ defines the mode or control state (e.g., controlling the elbow, hand, or fingers). This affects the target joint angle $\theta_{target}(t)$.

\subsubsection{Human Error Correction}
The human user perceives errors in the arm's movement and provides feedback: $e_h(t) = \theta_{desired}(t) - \theta_{actual}(t)$ where $\theta_{desired}(t)$ is the human's intended joint position, and $\theta_{actual}(t)$ is the actual joint position of the prosthetic arm. The user can provide a voice based feedback or adjust the EEG input accordingly to minimize this error.

\subsubsection{Prosthetic Arm Dynamics}
The joint angle $\theta(t)$ of the prosthetic arm is updated based on the predicted action $\hat{a}(t)$ and the human feedback $H(t)$: $\theta(t+1) = \theta(t) + K_a \cdot \hat{a}(t) + K_h \cdot H(t)$ where $K_a$ and $K_h$ are gain factors for the control action $\hat{a}(t)$ and the human feedback $H(t)$, respectively.

\subsubsection{Feedback Loop (Human-in-the-Loop)}
The complete control law that incorporates the human's real-time feedback is: $\theta(t+1) = \theta(t) + K_a \cdot f_{control}(S_{EEG}(t), W) + K_h \cdot \left( \theta_{desired}(t) - \theta_{actual}(t) \right)$. This equation represents the adjustment of the prosthetic joint angle based on both the classifier's output and the human’s corrective feedback. The values of $K_a$ and $K_h$ are determined experimentally, through trials where the system's performance is evaluated under different weights to find effective parameters to balance between automatic control and human feedback.

\begin{algorithm}
\footnotesize
  \caption{\small EEG-based Prosthetic Arm Control with HITL}\label{alg:BRAVE}
  \begin{algorithmic}[1]
    \Require EEG data stream, voice command input $V_c$, pretrained model weights for LSTM, CNN, RF, and human feedback module
    \Ensure Control signal to prosthetic arm motors
    \State Load trained LSTM, CNN, and RF models
    \State Set window size $W_s$ and step size $S_s$
    \State Connect to EEG data stream and voice input module
    \Statex \textbf{Stage 1: Data Collection \& Preprocessing}
    \While{EEG data available}
      \State Collect window of EEG data $W$ of size $W_s$
      \State Apply bandpass filtering and artifact removal
      \State Extract features $F_{\mathrm{RF}}$ for the Random Forest model
    \EndWhile
    \Statex \textbf{Stage 2: Inference with Ensemble Models}
    \State {LSTM}: Predict class $P_{\text{LSTM}} \gets \text{LSTM}(W_{\text{LSTM}})$
    \State {CNN}: Predict class $P_{\text{CNN}} \gets \text{CNN}(W_{\text{CNN}})$
    \State {RF}: Predict class $P_{\text{RF}} \gets \text{RF}(F_{\text{RF}})$
    
    \State Combine predictions $\mathbf{P} \gets [P_{\text{LSTM}}, P_{\text{CNN}}, P_{\text{RF}}]$
    \State Final prediction $P_{\text{final}} \gets \text{Meta-model}(\mathbf{P})$
    \Statex \textbf{Stage 3: Human-in-the-Loop Correction}
    \State $P_{\mathrm{corrected}} \gets \mathrm{HumanCorrection}(P_{\mathrm{final}})$
    \State Update the final decision: $P_{\text{final}} \gets P_{\text{corrected}}$
    \Statex \textbf{Stage 4: Command Generation \& Execution}
    \If{$V_c \neq \mathrm{None}$}
      \State Map voice command $V_c$ to control mode
    \EndIf
    \State Generate control signal based on $P_{\mathrm{final}}$ and mode
    \State Send control signal to prosthetic arm motors
  \end{algorithmic}
\end{algorithm}

\begin{figure}[ht]
    \centering
    \includegraphics[width=1\linewidth]{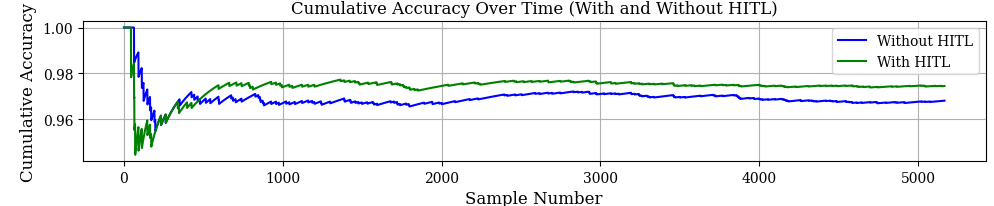}
    \caption{Cumulative accuracy over time for EEG signal classification with and without HITL. The plot illustrates how incorporating HITL corrections improves model accuracy as more samples are processed, leading to more consistent and accurate predictions compared to the non-HITL system.}
    \label{fig:HTIL}    
\end{figure}

%% file: Sections/conclusion.tex
\section{Conclusion}

The BRAVE system presents a novel {EEG-driven prosthetic control framework}, integrating {ensemble learning} and {voice-assisted mode switching} for {real-time, multi-DOF prosthetic control}. By combining {LSTM, CNN, and RF models} with {artifact rejection techniques} such as {ICA and CSP}, BRAVE achieves {high classification accuracy} while mitigating noise from EMG and EOG signals. 

With an average response latency of {150 ms}, the system ensures {low-latency actuation}, making it feasible for practical deployment. The {Human-in-the-Loop (HITL) correction} mechanism enhances adaptability, allowing real-time user adjustments for improved precision.

Future work will focus on {enhancing inter-subject generalization}, {optimizing for embedded systems}, and {expanding trials to diverse amputee populations}. BRAVE demonstrates significant potential in advancing {non-invasive brain-computer interfaces (BCIs) for prosthetic control}, offering an {accessible and adaptive} solution for next-generation assistive devices.

\section*{Acknowledgment}
This work was partially supported by the NYUAD Center for Artificial Intelligence and Robotics (CAIR), funded by Tamkeen under the NYUAD Research Institute Award CG010.